\documentclass[11pt,twoside]{article}
\usepackage{asp2004}
\usepackage{psfig}
\usepackage{epsf}
\usepackage{natbib}
\usepackage{graphics}
\usepackage{lscape}
\def\edcomment#1{\iffalse\marginpar{\raggedright\sl#1\/}\else\relax\fi}
\reversemarginpar

\begin{document}
\title{Evolution at very low and zero Z}
\author{Georges Meynet, Andr\'e Maeder and Sylvia Ekstr\"om}
\affil{Geneva Observatory, CH-1290 Sauverny, Switzerland}

\begin{abstract}
Rotation deeply affects the evolution of very metal poor massive stars. 
Indeed, even moderately rotating stars reach the break--up limit during
the Main--Sequence (MS) phase, they evolve rapidly to the red after the core H--burning phase
and important surface enrichment in CNO elements occurs at the supergiant stage. High rotation
and the self enrichment in heavy elements $Z$ tend
to enhance the quantity of mass lost by stellar winds. As a numerical example, we obtain that 
a 60 M$_\odot$ at $Z=0.00001$, with an initial rotation corresponding to 0.7 of the critical
velocity, loses half of its initial mass
during its lifetime. The ejected material, enriched in helium and CNO elements,
has interesting consequences for the early chemical evolution of galaxies.
\end{abstract}
\thispagestyle{plain}

\section{Introduction}

Pop III stars are not directly observed,
however they had an impact on the early photometric and chemical evolution of galaxies and they contributed
to the reionization of the Universe. Recently the interest in these objects has been stimulated 
by the observation of
extremely metal poor stars \citep*{Chr02} and the detection of galaxies at high redshifts \citep{Pe04}. 
Thus numerous grids for Pop III stars have recently appeared (\citeauthor{We00} \citeyear{We00}; \citeauthor{Chi02}
\citeyear{Chi02}; 
\citeauthor{He02} \citeyear{He02}; \citeauthor{Ma03} \citeyear{Ma03}; \citeauthor{No03} \citeyear{No03}; \citeauthor*{Tu03}
\citeyear{Tu03};
\citeauthor{Pi04} \citeyear{Pi04}).

As recalled above, for Pop III stars, there are no individual stars with which stellar models
can be compared and the check of the quality of the models remains quite indirect. An interesting approach is to use
for computing very metal poor stellar models the same physics which can successfully account for important observed properties  
of stars at higher metallicities. In this context it appears that rotation plays a key role in shaping many
of these properties (see Sect.~4 below). Thus, following the above approach,
it is quite natural to explore what are the effects of rotation at very low $Z$.
Recent attempts to explore these effects in metal poor environments have been performed by \citet*{Hea00} and \citet{Ma03}
at $Z$ = 0 and
by \citet{MMVIII} at $Z$ = 0.00001.
In this paper, after a recall of the main results obtained by non--rotating and rotating models at very low $Z$,
we explore, for the first time, the effects of very fast rotation on the evolution of massive, very metal poor stars,
resolving in a consistent way the equations for the transport of the angular momentum and of the
chemical species.

\section{Evolution of non--rotating very metal poor stars}

Evolutionary tracks for Pop III stars are shown together with solar metallicity models on Fig.~1.
At very low metallicity, the star formation processes may
lead to the formation of more massive stars that at high metallicity
(see Abel's contribution in the present volume). The evolutionary scenario of these very massive objects
have been recently discussed by \citet{He02}.
Among the most important differences between solar metallicity stars
and primordial stars, we can note the following ones (see the review by \citeauthor{Ch00} \citeyear{Ch00} for more details):

\begin{itemize}
\item Metal poor stars
are more compact than solar metallicity ones due to lower opacity. 
For instance, a 20 M$_\odot$ stellar model at Z = 0 has a radius on the ZAMS which is about a third
of the stellar radius of the corresponding model at solar metallicity. 
As a consequence, the moment of inertia of a star of a given initial mass is also smaller at lower metallicity:
typically it is equal to 0.62 10$^{57}$ [g cm$^2$] for a Pop III star and 5.72 10$^{57}$ [g cm$^2$]
for a solar metallicity star.
This may imply faster initial rotation rate, in case very metal poor stars
begin their evolution with the same amount of angular momentum than their more metal rich counterparts.

\begin{figure}[!t]
\plotfiddle{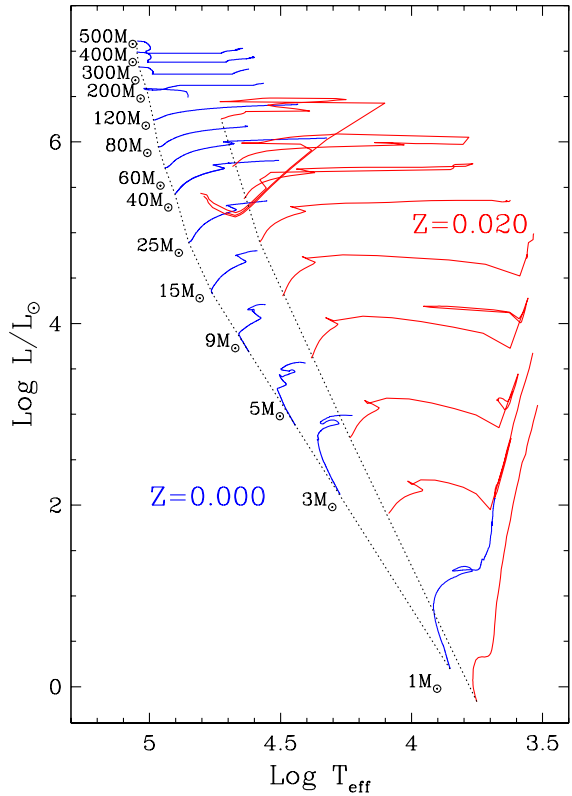}{10cm}{0}{40}{40}{-150}{-20}
\caption{Evolutionary tracks for non--rotating stellar models at solar and
zero metallicity. The computations were performed by \citet{Fei99}.}
\end{figure}

\item At solar metallicity, all stars with initial masses superior to $\sim$ 1.2 M$_\odot$ burn their hydrogen through the CNO cycle.
At $Z =0$ the lack of carbon for initiating the CN cycle implies that stars of all masses begin to burn their hydrogen through the pp chains. 
Since the energy generated by
these chains is not enough to support massive stars against gravity, the stars contract until the central temperatures reach sufficiently high value for
activating the $3\alpha$ reaction. When the abundance of carbon, synthesized by He--burning, reaches a value around 10$^{-9}$--10$^{-10}$ in mass fraction,
the CN cycle becomes fully efficient and the star recover the standard behavior.

\item According to the line--driven wind theory, 
the rate of mass loss depends on the metallicity of the layers of
the star where the absorption lines are formed \citep*{Ca75}. 
For the metallicity range between 10$^{-4}$ and 1 $\times$ Z$_\odot$, \citet{Ku02}
obtains that $\dot M(Z) \propto (Z/Z_\odot)^{0.5} \dot M(Z_\odot)$. 
Thus the mass loss rates decrease when the metallicity decreases.
Extrapolating the above scaling law for zero metallicity star would predict that these stars
lose no mass at all, at least until
some surface enrichment occurs modifying their metallicity.
However such a statement should be taken with caution. Other processes
which are metallicity independent may intervene for driving mass loss (e.g. effects of continuum radiation, pulsation).
At present, the intensity of the stellar winds at very low metallicity still
remains largely unknown.
\end{itemize}

\section{The variations of the effects of rotation with the metallicity}

Rotation induces numerous instabilities in the stellar interiors. These instabilities, at their turn, transport 
chemical species and angular momentum throughout the stellar interiors (\citeauthor{Hel00} \citeyear{Hel00}; \citeauthor{MMAA}
\citeyear{MMAA}). 
Rotation also affects the way massive stars lose mass through stellar winds \citep{MMVI}.
In the recent years the effects of rotation have been explored at different initial metallicities 
(\citeauthor{MMV} \citeyear{MMV}; \citeauthor{MMVII} \citeyear{MMVII}; \citeauthor{MMVIII} \citeyear{MMVIII}; 
\citeyear{MMX}; \citeyear{MMXI}).
Interestingly, these models show that given an initial mass and an initial equatorial velocity,
$\upsilon_{\rm ini}$, the lower the metallicity, the more efficient the chemical mixing and greater the chances for the star
to reach the break--up limit. Let us discuss these two points in more details.

\subsection{More efficient chemical mixing at low Z}

On panels ``a'' and ``b'' of Fig.~2, the variations with the radius
of the angular velocity in a solar metallicity and a very metal 
poor 20 M$_\odot$ model are compared. The quantity $X_c$ is the mass fraction of hydrogen
at the center. On can note three main differences between the two metallicities:
[1] The $Z=0.00001$ star
has a radius which is about half the radius of the solar metallicity model.
[2] $\Omega$ is higher in the $Z=0.00001$ model.
This comes from the fact that the two models began their evolution on the
ZAMS with the same value of the equatorial velocity. Since at $Z$ = 0.00001, the radius
is smaller, this implies that
$\Omega$ is greater ($\upsilon = \Omega R$).
[3] The gradients of $\Omega$ are much steeper in the lower metallicity
model. The explanation for this last difference arises from the fact that
the meridional velocity is
much smaller at lower metallicity\footnote{This mainly arises
from the Gratton-\"Opick term in the expression for the meridional circulation,
which scales as the inverse of the density  
(see \citeauthor{MMVII} \citeyear{MMVII} for a more detailed discussion).}. This can be seen from panels ``c'' and ``d'' of Fig.~2, which also
shows that during most of the Main-Sequence phase, the sign of U
(radial term of the vertical component of the meridional velocity) is negative, meaning an outward
transport of the angular momentum. Thus less angular momentum is transported outwards during the MS phase
and the gradients remain steeper. Contrary to angular momentum, the chemical species are more easily
transported in lower metallicity stars (see panels ``e'' and ``f''). This comes from the
fact that  the gradients
of $\Omega$ are steeper and thus the shear mixing is more efficient.
One notes also that in the lower metallicity model, 
the nitrogen enhancement
in the convective core is higher than in the solar metallicity model. Thus the gradient
of N abundance at the border of the core is also more important favouring its transport by shear mixing in the outer
radiative envelope. The nitrogen enhancement is more important at low $Z$, because the very metal poor models start with an
initial composition which is enriched in oxygen and slightly decreased in carbon and nitrogen ($\alpha$--enhanced
composition). This reflects the fact that early stellar generations were formed from material which was 
only enriched by massive star ejecta.

\begin{figure}[!t]
\plotfiddle{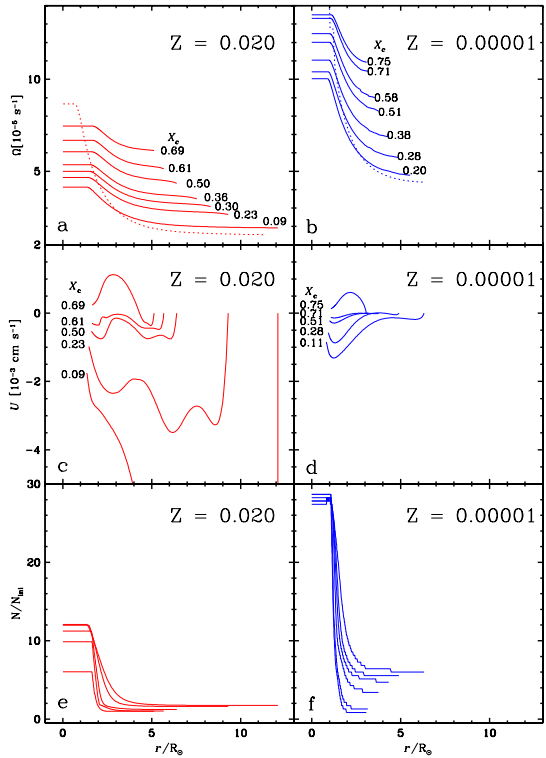}{10cm}{0}{40}{40}{-130}{-20}
\caption{Variation as a function of the radius in solar units of the angular velocity, $\Omega$, of the radial term
of the vertical component of the velocity of meridional circulation, $U$, and of the abundance of nitrogen
normalized to its initial value, $N/N_{\rm ini}$, during the Main sequence phase of 20 M$_\odot$ stellar models. 
In both cases $\upsilon_{\rm ini}$ = 300 km s$^{-1}$.
The left panels are for solar metallicity models, the right panels
are for $Z=0.00001$ models.
The values of $X_c$ are the mass fraction of hydrogen in the centre. The dotted lines in panels
``a'' and ``b'' show how $\Omega$ varies at the end of the Main--Sequence phase when the core contracts.
In panels ``e'' and ``f'', the models represented are the same
as those shown in panels ``c'' and ``d''. When time goes on, the surface becomes more and more enriched
in nitrogen, thus in these two panels, when time goes on, curves shift from bottom to top.}
\end{figure}

\subsection{Stars at low Z reach more easily the break--up limit}

As shown above, during most of the MS phase, angular momentum is transported outwards and thus
one expects that in absence of mass loss, the stellar outer layers will be accelerated. 
This is the main reason why at low metallicity, {\it i.e.} when the mass loss rates are low,
the surface velocity may increase with time during the MS phase. On the contrary, when the mass loss rates
are high, the outer layers are removed and the surface velocity decreases or at least is prevented from growing.
Polar winds might change this picture, although only for very high rotation rates
\citep{Ma99}.

In Fig.~3, the evolution 
of the ratio of the surface equatorial velocity to the break--up velocity\footnote{As shown by \citet{MMVI},
the expression of the break--up velocity is different for stars away or at proximity of the Eddington limit.
Near the Eddington limit, the critical velocity is smaller than $\upsilon_{\rm crit,1}$. Its expression is given
in the above reference. In the expression for the correcting factor of mass loss due to rotation, only
the ratio $\upsilon/\upsilon_{\rm crit,1}$ appears.} 
$\upsilon_{\rm crit,1}={2 \over 3}{GM \over R_{\rm pb}}$ is shown 
as a function of time for models at $Z$ = 0.020
and $Z$ = 0.00001.
Comparing the two plots for the two metallicities,
the most striking difference occurs for the models of 40 and 60 M$_\odot$ models. While
the solar metallicity models evolves away from the break--up limit, their lower metallicity counterparts
reach this limit at the end of their MS phases. It is interesting to remark here that the two
processes involved to explain such behaviors, namely the stellar winds and the outwards transport of the angular
momentum become {\bf both} less efficient when the metallicity (and/or
the initial mass) decreases. 
According to the present numerical models, clearly mass loss overcomes the transport effect
at solar metallicity, while the inverse situation occurs at $Z$ = 0.00001 at least for stars with initial masses 
between 40 and 60 M$_\odot$. It is likely that, for a given initial velocity, 
the range of initial masses which can reach the break--up limit
during the MS phase shifts to higher values when the metallicity decreases.

\begin{figure}[!t]
\plottwo{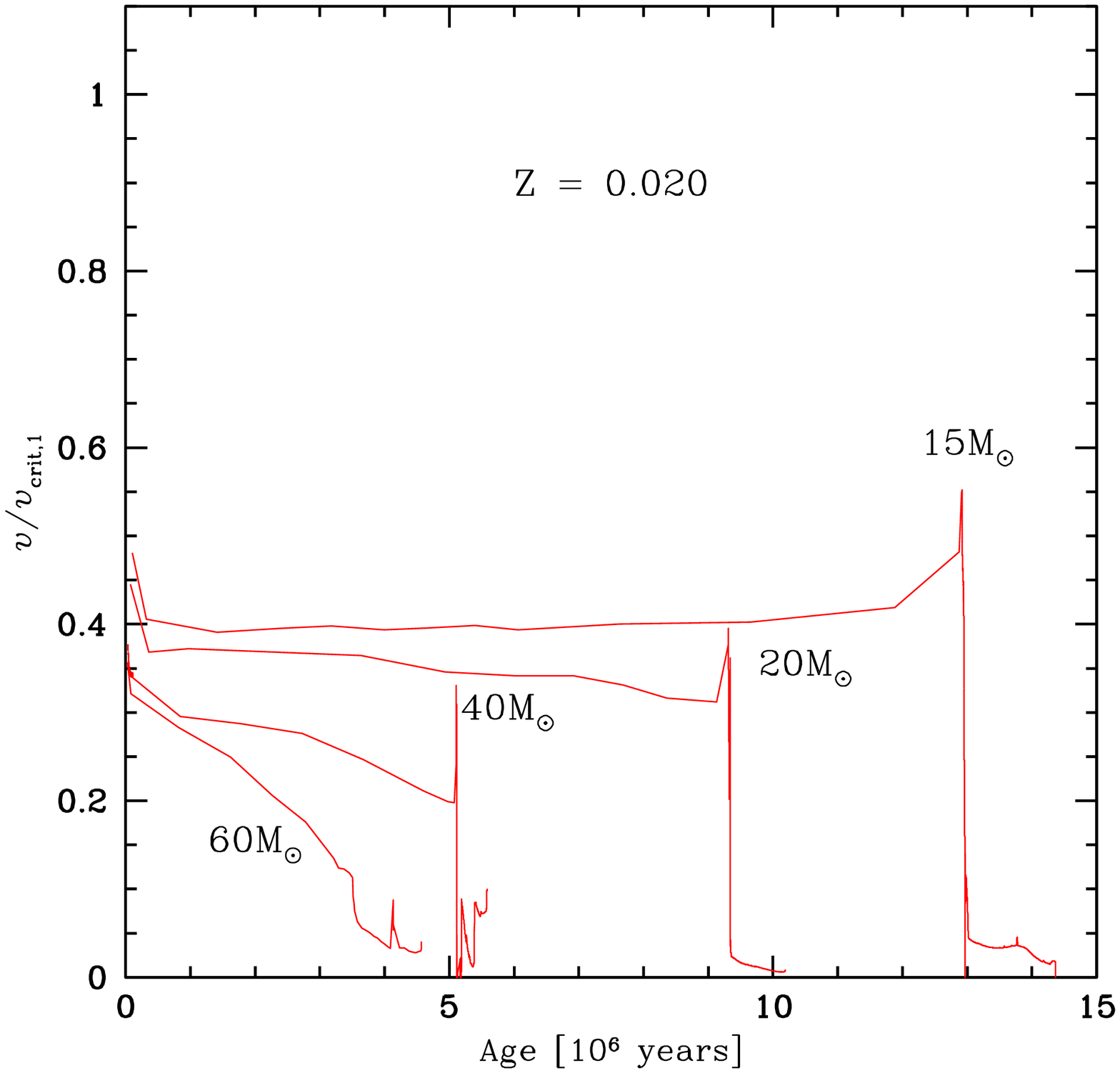}{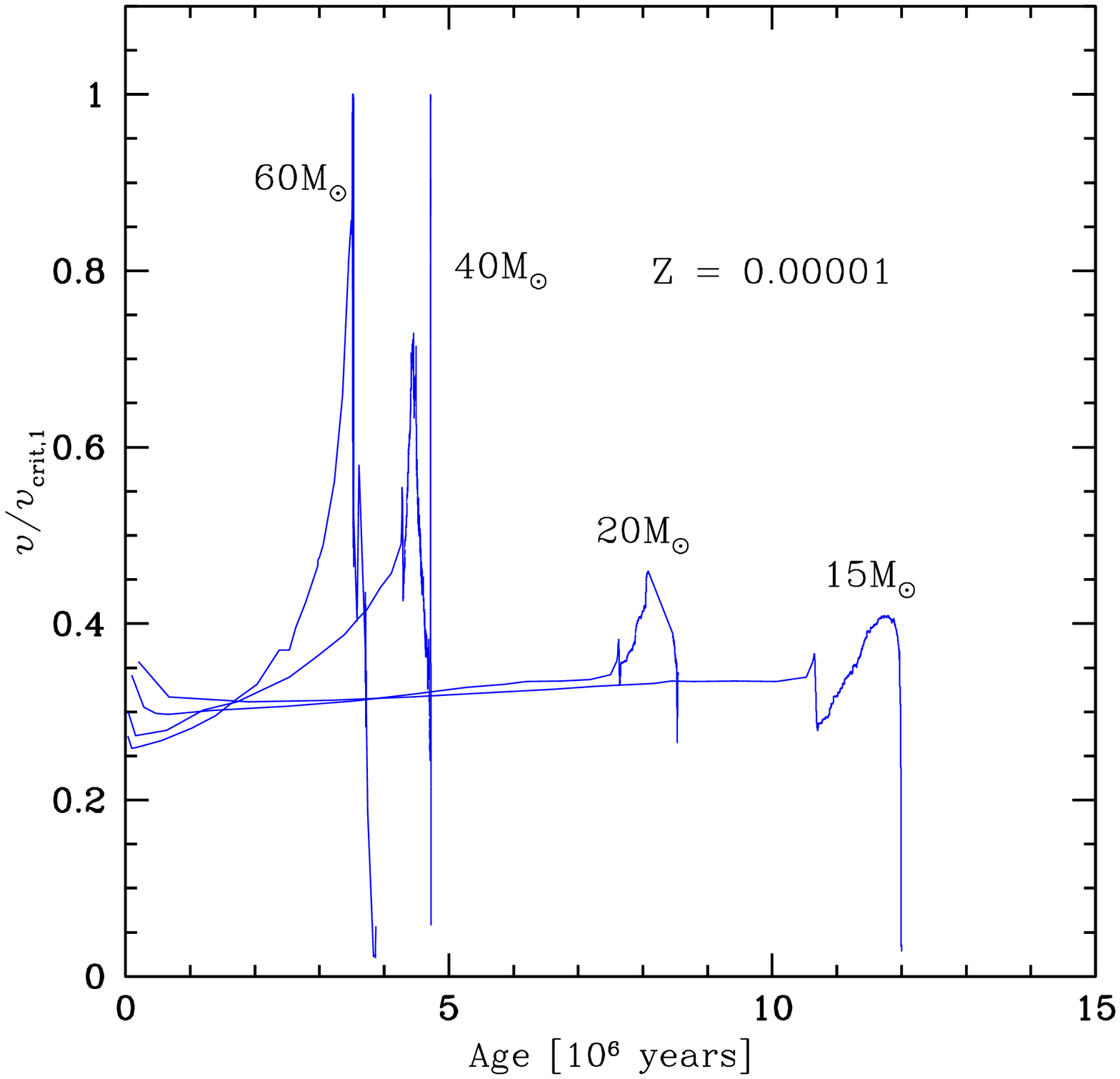}
\caption{Variation as a function of time, for two metallicities and various initial masses, 
of the ratio of the equatorial velocity at the surface
to the critical velocity $\upsilon_{\rm crit,1}={2 \over 3}{GM \over R_{\rm pb}}$, where $R_{\rm pb}$
is the polar radius at break--up. In all cases, the initial rotation on the ZAMS is $\upsilon_{\rm ini}$
= 300 km s$^{-1}$.}
\end{figure}

\section{Consequences for massive star populations at different metallicities}

The two effects discussed in the previous section have numerous observational consequences:
\begin{itemize}
\item For given the initial mass and rotational velocity, the lower the metallicity, the more 
important are the surface enrichments at a given evolutionary stage \footnote{Note that
exception to the above trend occurs for high mass stars ($M \ge ~ 40$ M$_\odot$) undergoing heavy mass loss rates
\citep{MMX}. In that case mass loss dominates the effects of rotation.}. The observations
by \citet{Ve04} of surface abundances of A--type supergiants in the Milky--Way
and in the SMC
support this view, although they might also be explained by a distribution of initial rotation
with more fast rotators at low $Z$.

\item For given the initial mass and rotational velocity, the lower the metallicity, the greater
the averaged surface velocity during the MS phase. The observed increase of the number ratio of Be stars
to B stars when the metallicity decreases (\citeauthor*{MG99} \citeyear{MG99}), also the recent rotational velocities
measured by \citet{Ke04} for early B--type stars in the LMC and the Galaxy gives some support to this view.
However, as above, the distribution of the initial velocity may be different at various metallicities
and may thus blur the evolutionary effect predict by the present stellar models.

\item Mixing induced by rotation favours the redwards evolution after the MS phase \citep{MMVII}. This effect
might explain the great number of red supergiants observed in the SMC.

\item The concomitant effects of rotation and mass loss by stellar winds play a key role in the process of Wolf--Rayet
star formation (\citeauthor{FL94} \citeyear{FL94}; \citeauthor{MMX} \citeyear{MMX}). It has been shown that models accounting for the effects of rotation
much better reproduce the variation with the metallicity of the number ratio of WR to O--type stars \citep{MMXI}.
Also the Z--dependance of the number ratio of type Ibc supernovae (supposed to arise from WR progenitors) to type II
supernovae is much better fitted by the rotating models.   
\end{itemize} 

In addition to these effects, rotation also
affects the stellar yields (see the contribution by Maeder in this volume). At low metallicity, it has been shown that
rotation naturally leads to the synthesis of primary nitrogen \citep{MMVIII}, 
in quantities which appear quite compatible with those
needed by chemical evolution models of galaxies for reproducing the behavior of the N/O ratio at
low Z (see e.g. \citeauthor{Pra03} \citeyear{Pra03}).

\section{Consequences for the evolution of very fast rotating metal poor stars}

Taking advantage of the good agreement between the stellar models and numerous observations 
for metallicities between 1/5 and 2 times solar (see the previous section),
it is interesting to explore the predictions of the models in  more extreme conditions. 
We shall focus our attention on a model of 60 M$_\odot$ star at $Z$ = 0.00001 
with an initial velocity equal to 800 km s$^{-1}$ (this corresponds to an initial value
of $\upsilon/\upsilon_{\rm crit,1} = 0.7$). There are three reasons for considering this case
of high initial rotation: 1) realistic simulations of the formation of the first stars in the Universe
show that these stars might begin their evolution with a high content of angular momentum \citep*{Ab02};
2) since the metal poor stars are more compact, a given value
of the angular momentum corresponds to a higher value of the initial equatorial velocity.
Typically,
at $Z = 0.020$, a value of $\Omega/\Omega_{\rm crit} = 0.5$ corresponds to a value of $\upsilon_{\rm ini}= 300$ km s$^{-1}$,
at $Z = 0$, the same value of $\Omega/\Omega_{\rm crit}$ corresponds to a value of $\upsilon_{\rm ini}= 800$ km s$^{-1}$.
3) There are some observational hints that
the distribution of initial rotation might contain more fast rotators at low $Z$ \citep{MG99};  

With respect to non--rotating stellar models, our very fast rotating model presents three main differences, which
all tend to enhance the quantities of mass lost by stellar winds:

1) The star remains near the break--up limit during about 2/3 of its MS lifetime. Due to this fact it loses about
6 M$_\odot$ which is much more than the 0.2 M$_\odot$ lost by the similar non--rotating model. 

2) After the MS phase, contrary to its non--rotating counterparts, which spends most of its
core He--burning phase in the blue region of the HR diagram, the model with $\upsilon_{\rm ini}$ = 800 km s$^{-1}$
rapidly evolves into cooler region of the HR diagram. More than 60\% of the core He--burning phase is spent near
log $T_{\rm eff}$ ~ 3.85, where the mass loss rates are higher than in the hotter part of the HR diagram.

3) As was obtained for less rapidly rotators \citep{MMVIII}, a bump of primary nitrogen appears at the position
of the H--burning shell in the stellar interior (see Fig.~4). This bump comes from the rotational diffusion of carbon and oxygen
produced in the He--core into the H--burning shell. Due to the redwards evolution mentioned just above,
a deep outer convective zone appears, which dredges--up CNO elements
to the surface. In the last computed model the mass fraction of CNO elements is more than
three orders of magnitude greater than the initial metallicity of the star ! 
This is well apparent on Fig.~4 which shows the evolution of the chemical
structure of the star at four evolutionary stages during the core He--burning phase. Their main characteristics 
are given in Table 1. One sees that
during the stages ``a'' to ``d'' the abundance in CNO elements at the surface increases
by a factor $~$1800. What are the effects of this important supply of CNO elements at the surface 
on the mass loss rates ?\footnote{Let us mention that during these stages the star remains more or less at the same position in the HR diagram
(log $L/L_\odot \sim$  6.08; log $T_{\rm eff} \sim$  3.86) and has a very low surface velocity (inferior to 5 km s$^{-1}$).}
The answer to the above question is not straightforward. 
The scaling law describing the metallicity dependance of the mass loss rates
involve the metallicity as a whole and describes the variation of $\dot M$ when all the heavy element abundances
are multiplied by the same factor, but do not say anything when  
the enhancement of the metallicity
is due to an increase of only CNO elements.
Moreover this law is based from studies using a solar mixture for the heavy elements and was obtained in the frame
of the line--driven wind theory, while
here we have a heavy elements mixture very different from the solar one and the star is in the supergiant stage
where other mechanisms than line-driven acceleration may trigger the mass loss rates.
The problem is thus intricate and certainly 
should be addressed through the study
of detailed stellar wind models for the evolutionary stage and the chemical composition considered here. 
For the present investigation, we decided to keep the usual variation of the mass loss rate with the metallicity,
{\it i.e.}  $\dot M(Z)= (Z/Z_\odot)^\alpha \dot M(Z_\odot)$, with $\alpha =0.5$ as devised by \citet{Ku02}. The  metallicity
$Z$ at the surface of the star may evolve due to internal mixing processes.
With such a rule, the arrival of important amounts of CNO elements at the surface boost the mass loss rate
at the supergiant stage. During the whole post--MS phase,
our stellar model  loses 23.5 M$_\odot$. Again, compared to the total mass lost by stellar winds by the
non--rotating model (0.4 M$_\odot$), this represents a huge enhancement, which completely changes the evolution of the star. 

\begin{table}[!t]
\caption{Characteristic of the models shown in Fig.~4.}
\smallskip
\begin{center}
{\small
\begin{tabular}{ccccccc}
\tableline
\noalign{\smallskip}
Stage & Age & Mass & $\dot M$ & C & N & O\\
      & [Myr.] & [M$_\odot$] & [$\dot M$ in M$_\odot$ yr$^{-1}$] & & &       \\
\noalign{\smallskip}
\tableline
\noalign{\smallskip}
a & 3.43 & 51.58 & -4.835 & $<$ 5 10$^{-7}$ & 0.000004 & 0.000003\\
b & 3.65 & 46.88 & -4.023 & 0.000055        & 0.000420 & 0.000153\\
c & 3.70 & 40.06 & -3.676 & 0.000602        & 0.001863 & 0.001305\\
d & 3.73 & 30.70 & -3.478 & 0.002142        & 0.005719 & 0.004903\\
\noalign{\smallskip}
\tableline
\end{tabular}
}
\end{center}
\end{table}

\begin{figure}[!t]
\plotfiddle{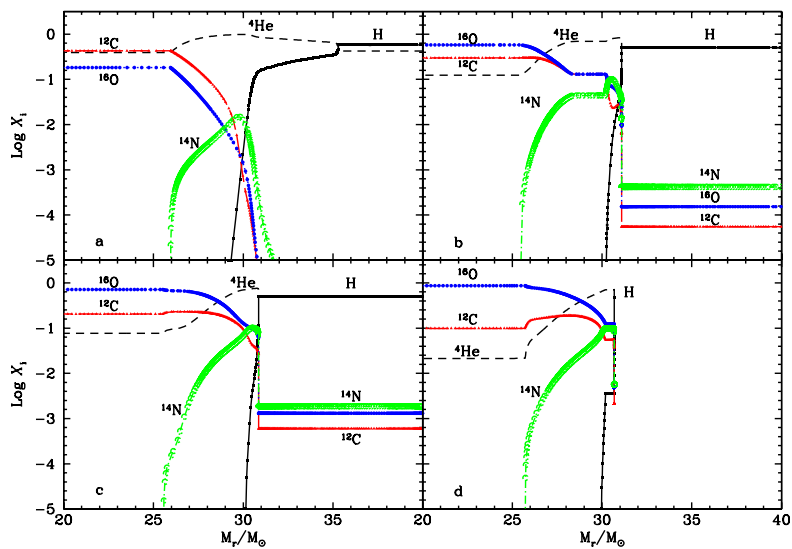}{7cm}{-90}{40}{40}{-180}{230}
\caption{Variation as a function of the lagrangian mass of the mass fraction of
various elements at four stages during the core He--burning phase of the
60 M$_\odot$ model at $Z$ = 0.00001 with $\upsilon_{\rm ini}$ = 800 km s$^{-1}$. Panels
``a'' to ``d'' correspond to the evolutionary stages ``a'' to ``d'' described in Table 1.}
\end{figure}

\begin{figure}[!t]
\plotfiddle{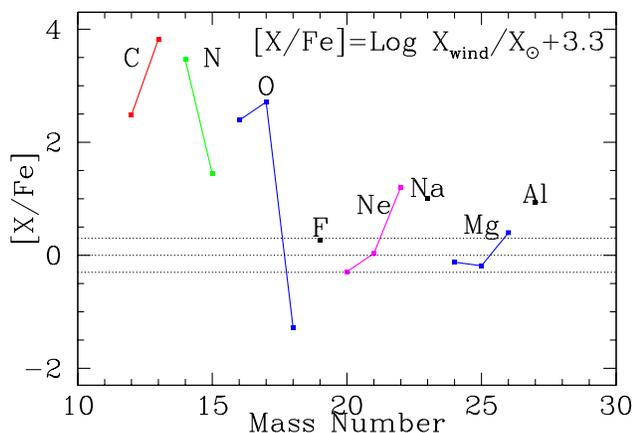}{5cm}{-90}{30}{30}{-160}{170}
\caption{Chemical composition of the matter expelled through stellar winds during the lifetime
of a 60 M$_\odot$ at $Z$ = 0.00001 with $\upsilon_{\rm ini}$= 800 km s$^{-1}$. Note that
3.3 = log ($Z_\odot/Z$). The horizontal dotted lines show the
solar ratios (middle) and a factor two enhancement/depletion.}
\end{figure}

This would have an important impact on the early chemical evolution of galaxies. 
For instance,
our 60 M$_\odot$ model ends with a
30 M$_\odot$ CO core and likely will form a Black--Hole at the end
its evolution. If no ejection occurs during this final stage, then such a star would only contribute
through its winds to the enrichment of the interstellar medium. 
The chemical composition of the wind ejecta is shown in Fig.~5. It presents important overabundances of carbon, nitrogen and oxygen.
Interestingly,
the abundance patterns obtained present some similarities with those
observed at the surface
of carbon--rich, extremely metal poor halo stars
(see e.g the measured abundances for the Christlieb'star, \citeyear{Chr02}).
Also new synthesized helium is ejected in important quantities (nearly 6 M$_\odot$ in the wind), which might have interesting
consequences when primordial helium is deduced from the observation of very metal poor regions.
Of course such a fast rotating star might still eject some material at the time of the
supernova collapse. However, the computation of our model was stopped at the end of the core He--burning phase,
thus, at a too early stage for giving useful predictions of the supernova yields.
Works are in progress in order to obtain such data.

If indeed very metal poor stars should be fast rotators, then this would change
considerably our views on the way these metal poor stars evolve and affect the early
chemical evolution of galaxies. More informations on the way
the distribution of rotation varies as a function of metallicity, as well as on the behavior of the mass loss rates
at very low metallicity are needed in order to obtain a better view of the first stellar generations in the
Universe.

\end{document}